\documentclass[aps,prd,twocolumn,superscriptaddress]{revtex4-1}

\usepackage{graphicx}
\usepackage{color}
\usepackage{amsfonts, amsmath}

\newif\ifoneauthor
\oneauthortrue

\newcommand{\e}[0]{{\rm e}}

\usepackage[colorinlistoftodos]{todonotes}
\usepackage[colorlinks=true, allcolors=blue]{hyperref}
\usepackage[printwatermark]{xwatermark}

\begin{document}

\title{Sensing Optical Cavity Mismatch with a Mode-Converter and Quadrant Photodiode}


\author{Fabian Maga\~{n}a-Sandoval}
\email[]{fabian.magana-sandoval@ligo.org}
\affiliation{Department of Physics, Syracuse University, NY 13244, USA}
\author{Thomas Vo}
\email[]{thomas.vo@ligo.org}
\affiliation{Department of Physics, Syracuse University, NY 13244, USA}
\author{Daniel Vander-Hyde}
\email[]{daniel.vander-hyde@ligo.org}
\affiliation{Department of Physics, Syracuse University, NY 13244, USA}
\author{J. R. Sanders}
\email[]{jax.sanders@marquette.edu}
\affiliation{Department of Physics, Syracuse University, NY 13244, USA}
\author{Stefan W. Ballmer}
\email[]{stefan.ballmer@ligo.org}
\affiliation{Department of Physics, Syracuse University, NY 13244, USA}


\date{\today}

\begin{abstract}
We present a new technique for sensing optical cavity mode mismatch and alignment by using a cylindrical lens mode converting telescope, radio-frequency quadrant photodiodes, and a heterodyne detection scheme. The telescope allows the conversion of the Laguerre-Gauss bullseye mode ($LG_{01}$) into the $45^{\circ}$ rotated Hermite-Gauss (``pringle") mode ($HG_{11}$), which can be easily measured with quadrant photodiodes. We show that we can convert to the $HG$ basis optically, measure mode mismatched and alignment signals using widely produced radio-frequency quadrant photodiodes, and obtain a feedback error signal with heterodyne detection.
\end{abstract}


\maketitle

\section{Introduction}

Optical cavities are ubiquitously used in interferometry and in particular in the Laser Interferometer Gravitational-wave Observatory (LIGO). Optical cavities must be aligned and mode matched to yield the best performance. Alignment hardware and schemes are well developed \cite{Morrison:94} while mode matching hardware and schemes have not attained the same level of maturity. This leads to a reduction of sensitivity for gravitational-wave detectors such as Advanced LIGO \cite{AdvancedLIGO2015}. Monitoring mode matching and dynamically correcting for it will ensure the best performance of future Advanced LIGO upgrades. This is particularly true for the use of non-classical squeezed vacuum states of light \cite{oelker2014squeezed} currently being commissioned for use in Advanced LIGO, as these states are exponentially sensitive to any optical loss mechanism, including imperfect mode matching.
 
A theoretical description of misalignment and mode mismatch is done by Anderson \cite{anderson1984alignment}.
Optical cavity misalignment and mode mismatching generate higher order optical modes. The first relevant modes for cavity misalignment are the well-known Hermite-Gaussian modes $HG_{10}$ and $HG_{01}$, while the dominant mode relevant for mode-mismatch is the Laguerre-Gaussian $LG_{01}$ mode ($LG_{lp}$ where $l$ is the azimuthal mode index and $p$ is the radial mode index).
Higher order mode-sensing techniques currently utilize CCD cameras, clipped photodiode arrays \cite{miller2014length}, or bullseye photodiodes (BPD) \cite{mueller2000determination}. These sensors provide feedback error signals for correcting either the beam waist size or waist location, but also have drawbacks. Some of the drawbacks include slow signal acquisition for CCD sensors, 50\% reduction in sensing capabilities for clipped arrays, and expensive custom parts that are difficult to setup for bullseye photodiodes. 

\begin{figure}[h]
\label{fig:Modeconverter}
\centering
\includegraphics[width=.4\textwidth]{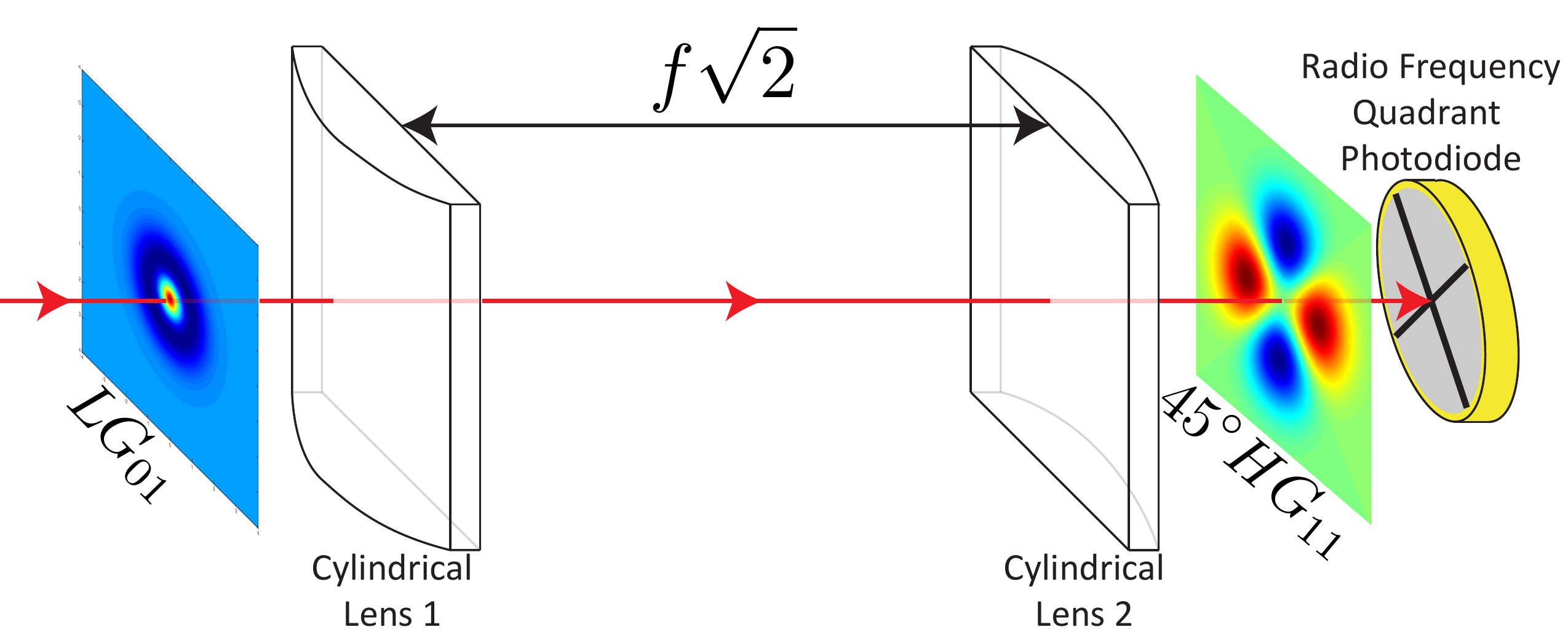}
\caption{\protect The concept of sensing mode mismatch using a mode converter and heterodyne detection on a quadrant photodiode. A $LG_{01}$ mode is converted to a $45^{\circ}$-rotated $HG_{11}$ mode with a $\frac{\pi}{2}$ mode converter. The mode converter consists of two cylindrical lenses spaced by $f\sqrt{2}$ where $f$ is focal length. The incoming beam waist size $w_0$ is centered between the two cylindrical lenses and related to the cylindrical focal length via $f(w_{0}) = \frac{\pi w_0^2}{\lambda}/{(1+\frac{1}{\sqrt{2}})}$ \cite{o2000mode}.}
\end{figure}

While sensing mode matching is challenging, alignment sensing is well developed in comparison and relies on easily available RF quadrant photodiodes.
By applying a $\frac{\pi}{2}$ mode converter \cite{o2000mode}, we show that the $LG_{01}$ mode turns into a $45^{\circ}$-rotated $HG_{11}$ mode, shaped perfectly for a quadrant photodiode. After sensing with a quadrant photodiode (QPD) we are free to use well-known heterodyne detection methods \cite{anderson1984alignment, black2001introduction, miller2014length, mueller2000determination} to extract a robust mode matching error signal.
Thus the mode converter allows using the usually discarded ``pringle" quadrant combination (+-+-) in existing alignment schemes for mode-matching feed-back (see FIG. \ref{fig:Modeconverter}).
This sensing scheme remains valid for large deviations from ideal mode-matching where a number of higher order modes contribute to the error signal (appendix \ref{appSG}).

\section{Modeling Mode Conversion and Error Signals}

\subsection{Mode Converter}
\begin{figure}[h]
\centering
\includegraphics[width=.4\textwidth]{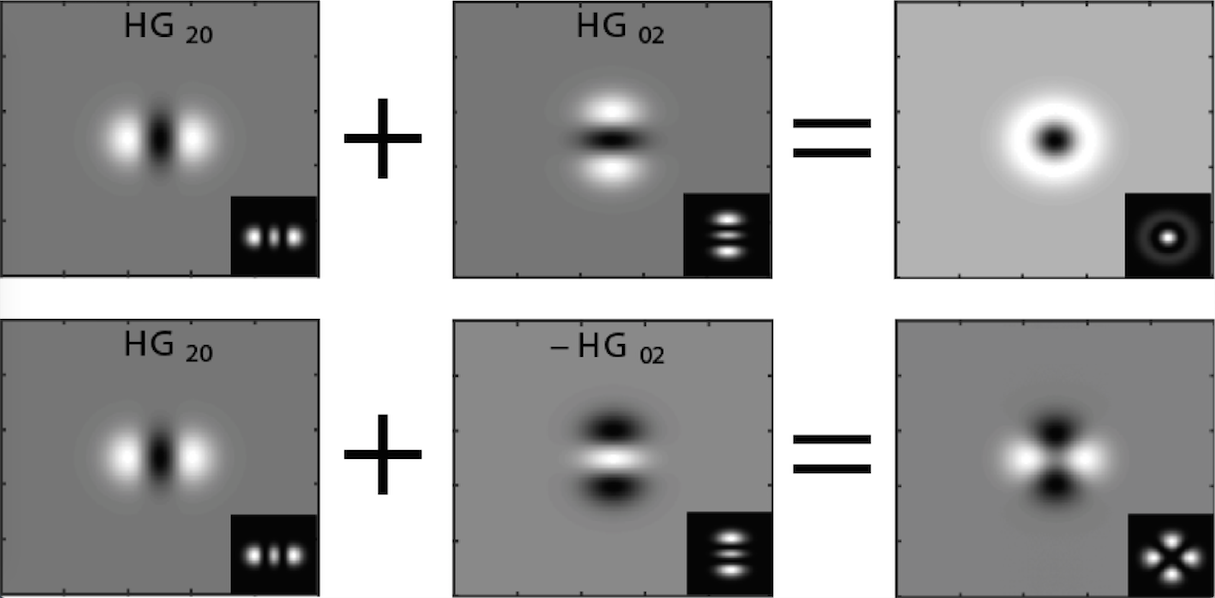}
\caption{\label{fig:LG01beamdecomposition} Beam decomposition of the $|LG_{01}\rangle$ and $|HG_{11}^{45^{\circ}{\rm rot}}\rangle$ mode in the $HG$ basis. Shown are the (real) field amplitudes of the two relevant HG modes (left) and the resulting modes (right). The intensity profile for each field is plotted in the bottom right corner for each field image.
}
\end{figure}

To understand how we can convert a Laguerre-Gauss $|LG_{01}\rangle$ mode into a $45^{\circ}$ rotated Hermite-Gauss $|HG_{11}\rangle$ mode we can decompose the beam in the $|HG_{nm}\rangle$ basis.
The $|LG_{01}\rangle$ bullseye mode is the sum of exactly two modes
\begin{equation}
\label{eqLGdef0}
|LG_{01}\rangle =\frac{1}{\sqrt{2}}|HG_{20} \rangle + \frac{1}{\sqrt{2}}|HG_{02} \rangle,
\end{equation}
as illustrated in FIG. \ref{fig:LG01beamdecomposition}. 
However, if we instead subtract the $HG$ components instead of adding them, we will find that
\begin{equation}
\label{eqHGrotdef0}
|HG_{11}^{45^{\circ}{\rm rot}}\rangle =\frac{1}{\sqrt{2}}|HG_{20} \rangle -\frac{1}{\sqrt{2}}|HG_{02} \rangle.
\end{equation}
where $|HG_{11}^{45^{\circ}{\rm rot}}\rangle$ is the $45^{\circ}$ rotated $|HG_{11}\rangle$ mode. 

This reveals that the only difference between a $|HG_{11}^{45^{\circ}{\rm rot}}\rangle$ mode and a $|LG_{01}\rangle$ mode is a sign flip along one axis, converting a parabolic wave front into a hyperbolic saddle point wave front.

\begin{figure}[h]
\centering
\includegraphics[width=.5\textwidth]{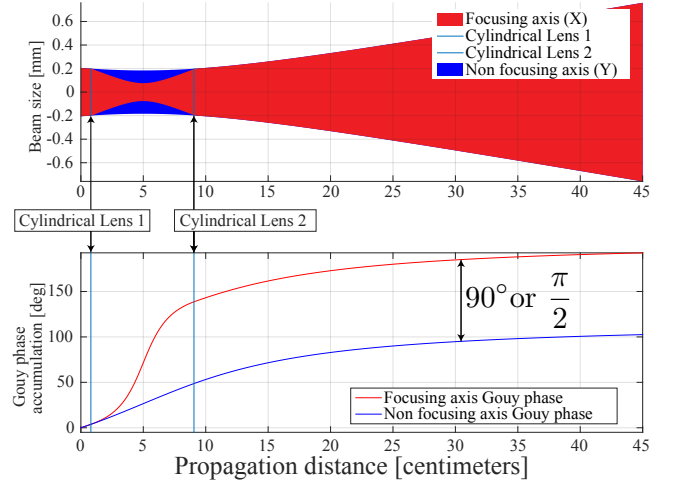}
\caption{\label{fig:ModeConverterGouyPhaseAccumulation} A cylindrical lens mode converter is shown.
The beam shape is plotted with respect to the cylindrical lens focusing and non focusing axis.
As the beam passes through the mode converter a factor of $90^{\circ} \mbox{ or }\frac{\pi}{2}$ is accumulated in the focusing axis while the non focusing axis experiences normal Gouy phase accumulation.
}
\end{figure}

A $\frac{\pi}{2}$ mode converter creates a region where Gouy phase is accumulated at different rates for the each transverse axis as seen in FIG. \ref{fig:ModeConverterGouyPhaseAccumulation}. 
The cylindrical lens focusing axis accumulates $\frac{\pi}{2}$ more phase than the non focusing axis.
Since second order modes accumulate twice the Gouy phase, the
$|HG_{20}\rangle$ and the $|HG_{02}\rangle$ see a phase accumulation difference of exactly $\pi$.
This flips the sign along one axis via the Euler identity, $-1=e^{i\pi}$, and creates the desired effect seen in FIG. \ref{fig:LG01beamdecomposition} and FIG. \ref{fig:ALLmodeconversions}. 
Designing a mode converter is described in appendix \ref{DesignofMC} and by Beijersbergen \cite{beijersbergen1993astigmatic}.


\subsection{Mode-Match Error Signal}
\label{mmsensors}
\begin{figure}[h]
\centering
\includegraphics[width=.5\textwidth]{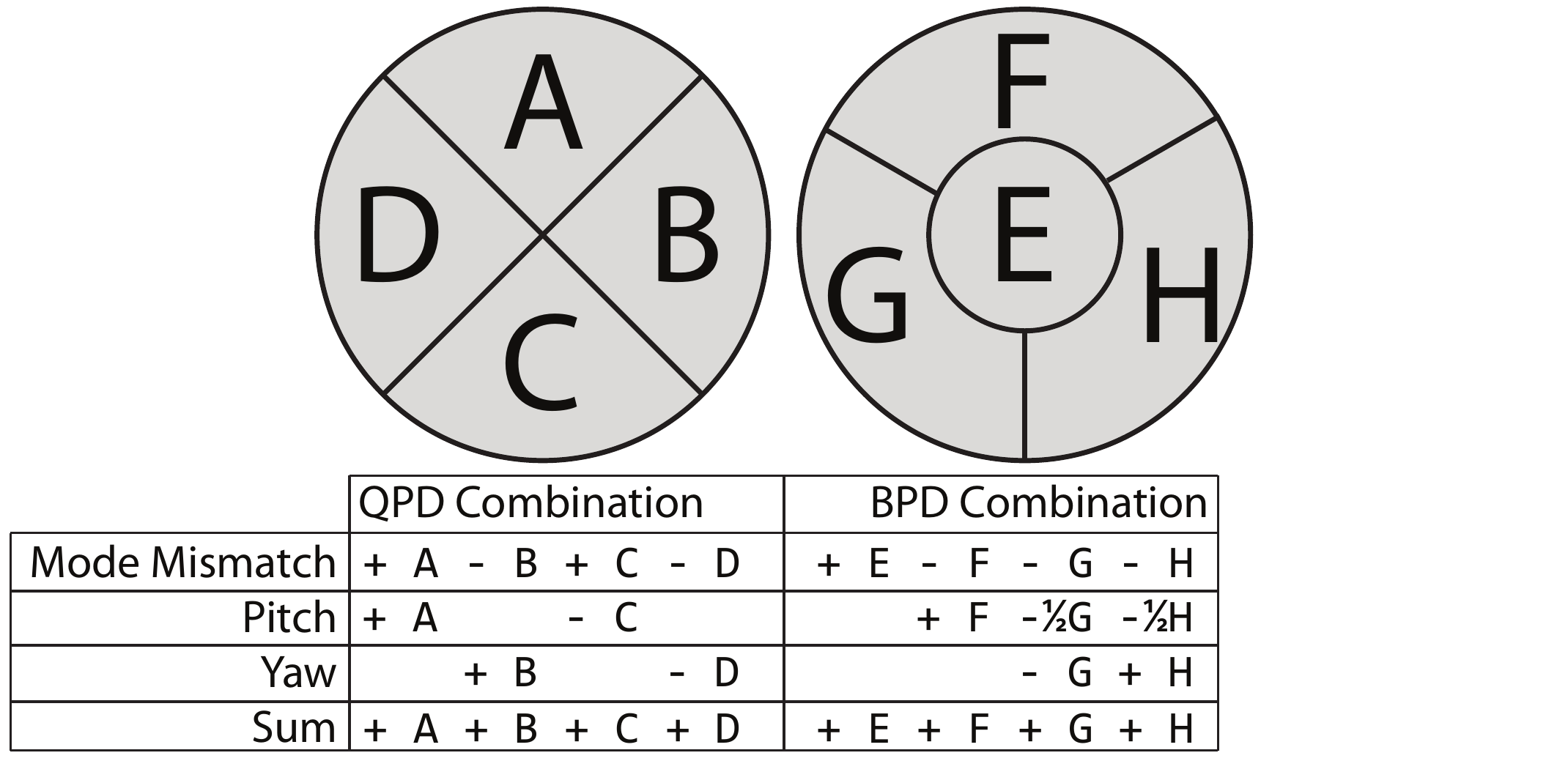}
\caption{\label{fig:Quadrant_Photodiode} Combining the photodiode segments can yield pitch, yaw, sum, and beam waist size or position. Left: quadrant photodiode (QPD). Right: bullseye phododiode (BPD).}
\end{figure}

As with any Pound-Drever-Hall-style sensing \cite{anderson1984alignment,Morrison:94,miller2014length,mueller2000determination} scheme we sense the light using RF-demodulated photodiodes. Since, after passing the mode-converter, the mode-matching information is contained in the $|HG_{11}^{45^{\circ}{\rm rot}}\rangle$ mode, we use a quadrant photodiode rotated by a $45^{\circ}$ relative to the mode-converter cylindrical lens axis. After demodulation we add the diagonals and subtract them from each other to get the error signal, see FIG. \ref{fig:Quadrant_Photodiode}. In contrast, for a bullseye photodiode-based scheme we take the inner segment subtracted by the sum of the outer segments. Both schemes also allow sensing alignment and length signals (pitch, yaw and length).

\subsection{Maintaining alignment sensing}
\label{multisensors}
Typical optical cavity alignment sensing requires the ability to measure the $|HG_{01}\rangle$ and $|HG_{10}\rangle$ modes with a quadrant photodiode \cite{anderson1984alignment}. 
Thus we need to examine what happens to the modes generated by misalignment after they pass through the mode converter. If the nodal lines of the $|HG_{01}\rangle$ and $|HG_{10}\rangle$ modes are at $0^{\circ}$ or $90^{\circ}$ relative to the mode-converter cylindrical lens axis, the two modes are passed unchanged - albeit with a relative phase shift of $\frac{\pi}{2}$ between the two.
In a Pound-Drever-Hall-style alignment sensing scheme these modes beat against the fundamental $|HG_{00}\rangle$ mode, which passes the mode-converter unchanged (see equations \ref{eqnHGdef1} to \ref{eqnHdef}). Thus all alignment signals are still present, but the signals for one axis are shifted by $\frac{\pi}{2}$ in sensing Gouy phase relative to the other axis.

The standard alignment sensing unit in gravitational-wave detectors is a pair of RF quadrant photodiodes (wave front sensors), installed on a sled behind a beam splitter and a Gouy phase telescope, guaranteeing that the two photodiodes are $90^{\circ}$ Gouy phase apart. This setup guarantees that every possible signal is accessible. Installing a mode-converter in front of such a sensing sled would thus preserve all alignment signals, only requiring a new sensing matrix. At the same time it would provide a sensor for one of the two possible mode-matching degrees of freedom. Since the orthogonal mode-matching degrees of freedom is separated by $45^{\circ}$ apart in Gouy phase, it would not be sensed. 

If sensitivity to only one mode-matching degrees of freedom is required - as is often the case in gravitational-wave interferometer applications - this simple upgrade would suffice, as long as the Gouy phase of the diodes is carefully chosen. If sensing of both mode-matching degrees of freedom is required, one can compromise by placing the second photodiode at an intermediate Gouy phase,  somewhere between $45^{\circ}$ and $90^{\circ}$ Gouy phase away from the first photodiode. The optimal location depends on the sensing noise requirements. Alternatively one can choose to install a third RF-quadrant photodiode $45^{\circ}$ Gouy phase away from diodes one and two.

\subsection{Controlling mode-match in interferometers}
Besides optimizing optical gain, the quality of mode-matching between the various optical cavities in a gravitational-wave interferometer matters for two critical reasons. First, gravitational-wave interferometers like Advanced LIGO are now routinely using squeezed vacuum injected from the anti-symmetric port to reduce the quantum noise level \cite{oelker2014squeezed}. Imperfect mode-matching couples the regular quantum vacuum fluctuations back into the readout, reducing the benefit from using squeezed vacuum. Second, a number of important noise sources, such as for example intensity noise on carrier and sideband, phase noise on the sideband and beam jitter, couple to the gravitational-wave readout through higher-order modes in the interferometer. While both 1st and 2nd order modes are problematic, an alignment system actively cancels 1st-order modes. Thus the largest higher-order modes are typically 2nd-order; they dominate the noise couplings unless an active mode-matching system suppresses them.

Alignment control of gravitational-wave interferometers has been extensively studied \cite{Morrison:94}, \cite{Mavalvala97}, \cite{Hirose:2009me}, \cite{Dooley:13}. All systems are an extension of the single-cavity Pound-Drever-Hall control scheme. The key differences when going to a more complicated system of coupled cavities are: (i) The beam splitter changes the sensing basis from individual arm cavities to common/differential arm cavities, sensed at the symmetric and anti-symmetric port of the beam splitter. And (ii) alignment signals from mirrors in coupled cavities can be disentangled by using multiple sensors operating with optical sidebands that are resonant in different portions of the coupled cavities. This design philosophy directly translates to mode-matching sensors, except that the system uses the second order transverse modes instead of the first order ones, requiring the type of sensors described in paragraphs \ref{mmsensors} and \ref{multisensors}.

\section{Experimental Demonstration}

\subsection{Experimental Layout}

\begin{figure}[h]
\centering
\includegraphics[width=.4\textwidth]{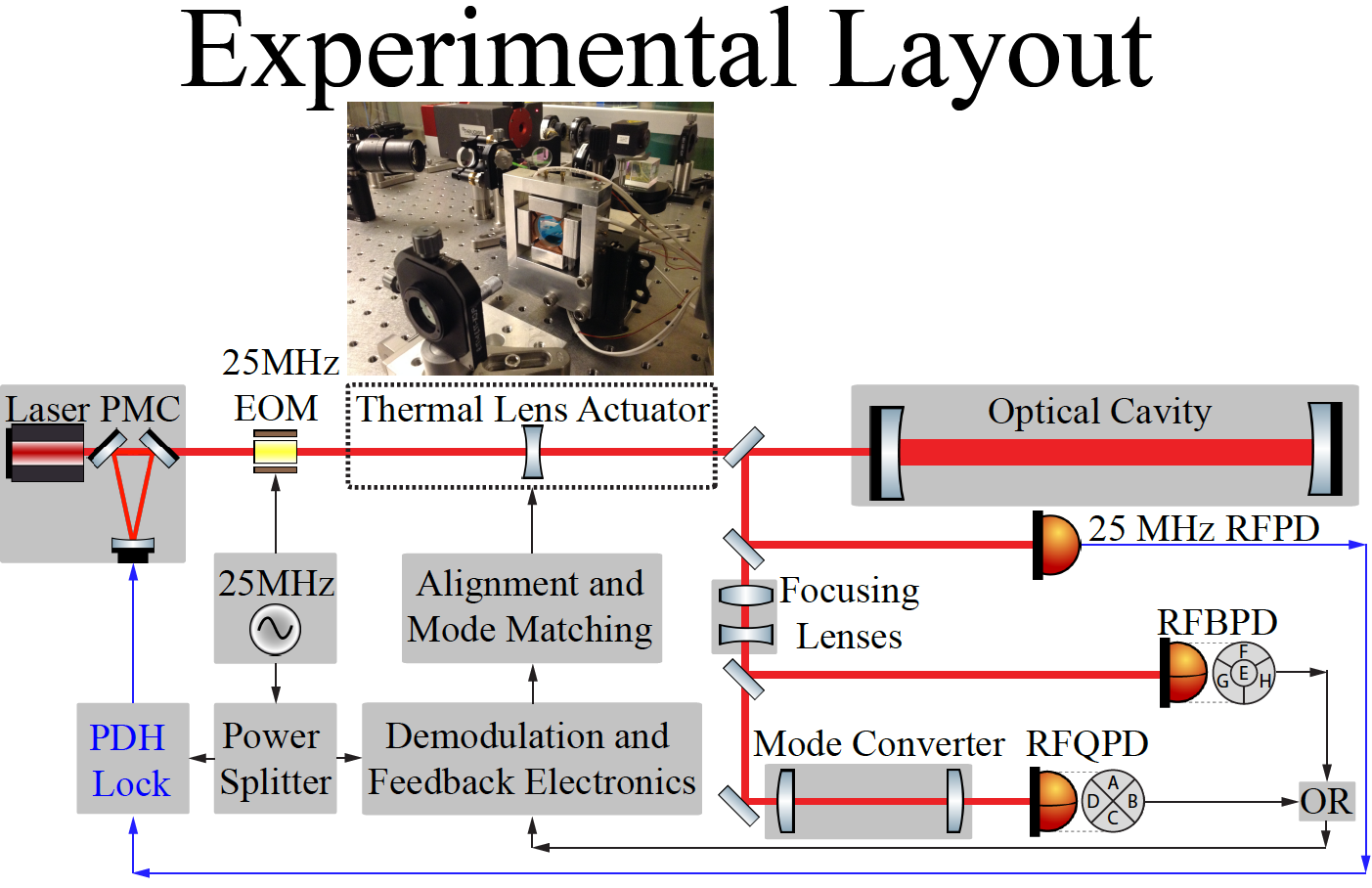}
\caption{\label{fig:ExperimentalSetup} A locked pre-mode-cleaner produces a beam that enters a 25 MHz EOM then propagates to a four segment thermal lens actuator. The thermal lens allows for pitch, yaw and beam size control. The cavity is aligned and well mode-matched to the beam. The cavity reflected beam is split into three paths. The first path leads to a single segment Pound-Drever-Hall (PDH) locking photodiode. The second and third path lead to a Gouy phase telescope that also shapes the beam for the $\tfrac{\pi}{2}$ mode converter. The quadrant photodiode (RFQPD) is in the path with the mode converter while the bullseye photodiode (RFBPD) is not. The Gouy phase at both bullseye photodiode and quadrant photodiode are similar. After demodulation, the signals are combined in the data acquisition system. Pitch, yaw, sum and mode-matching error signals are extracted. The cavity reflected power is attenuated by a factor of 0.30 before reaching the quadrant photodiode and 0.12 before reaching the bullseye photodiode.}
\end{figure}

The adaptive mode matching experiment at Syracuse University was built to study and provide mode matching sensor solutions for Advanced LIGO. 
FIG. \ref{fig:ExperimentalSetup} shows the optical layout we used to compare two types of wavefront sensing photodiodes.

A 1064 nanometer wave length Nd:YAG Mephisto S laser beam passes through a 13 MHz locked triangular mode cleaner. 
The triangular mode cleaner feedback and sensing electronics are not shown, but consist of a typical Pound-Drever-Hall (PDH) loop. 
The beam then passes through a 25 MHz EOM for PDH locking and wave front sensing. 
The phase modulated beam propagates to mode matching lenses and then to a four segment thermal lens actuator 
\cite{arain2010adaptive, liu2013feedback, AOEengineeringSpecs}.
A telescope is built around the thermal lens actuator such that the beam spot size is as big as possible without clipping on the 1 inch optic.
The beam then enters a well-aligned and mode-matched optical cavity.
The reflected beam continues through a Gouy phase telescope that also mode matches to a cylindrical lens mode converting telescope. 
Additionally, this telescope ensures that the beam size at the bullseye photodiode has the correct size and Gouy phase. 
A radio-frequency bullseye photodiode (BPD) and a radio-frequency quadrant photodiode (QPD) are placed at similar Gouy phases for a sensing comparison. 
The cavity reflected power is attenuated by a factor of 0.30 and 0.12 on the QPD and BPD respectively by various beam splitters.
The optical power is then sensed, demodulated, and sent to a digital data acquisition system.
In the digital system, the signals of each segment can then be combined to produce error signals.


\begin{figure}[h]
\centering
\includegraphics[width=.44\textwidth]{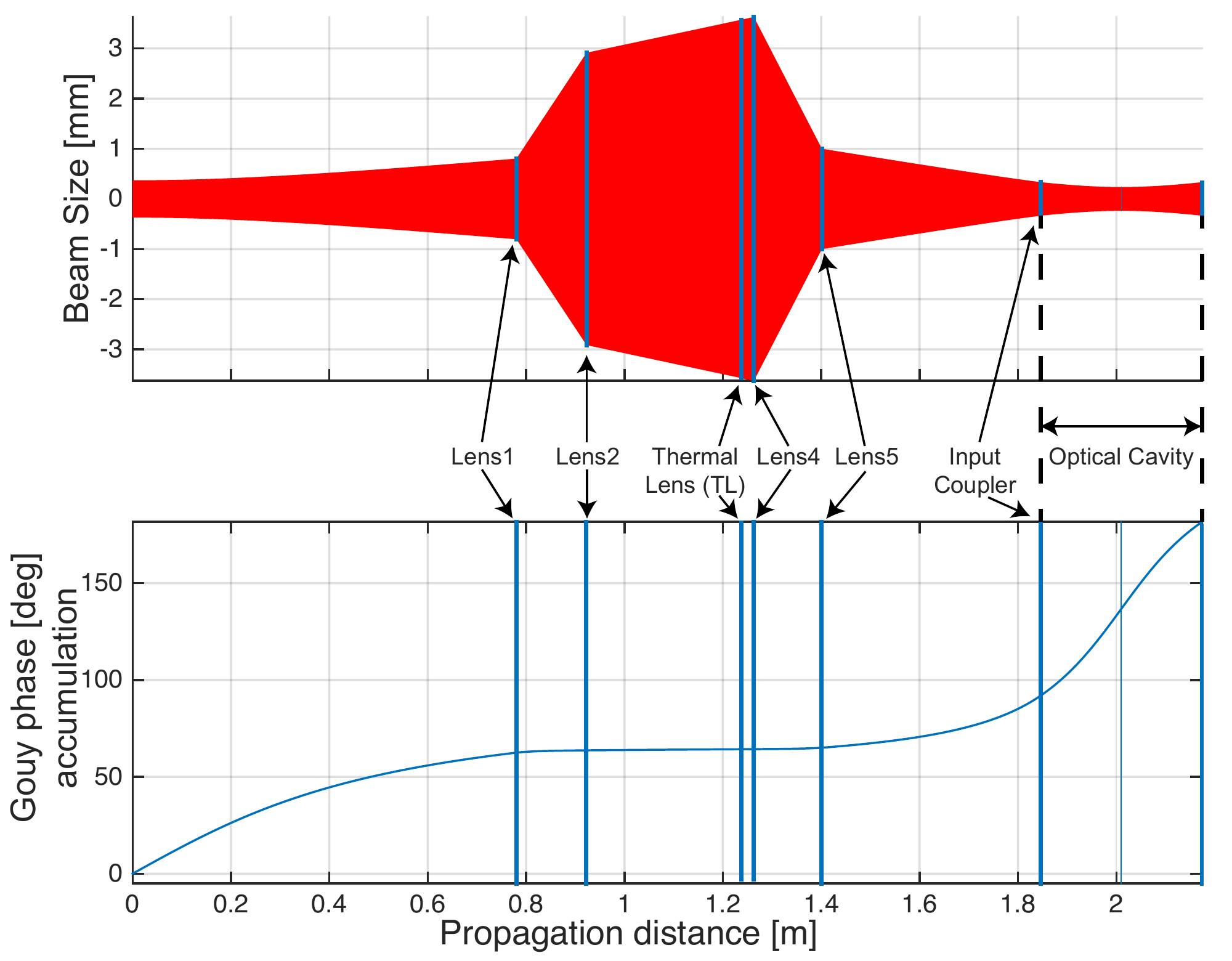}
\caption{\label{fig:TelescopeToCav} The telescope design from the mode cleaner to the two mirror cavity is shown. Beam size and Gouy phase accumulation are plotted against propagation distance. Lenses and the cavity waist location are represented by vertical lines further described in Table 1. }
\end{figure}

\begin{figure}[h]
\centering
\includegraphics[width=.48\textwidth]{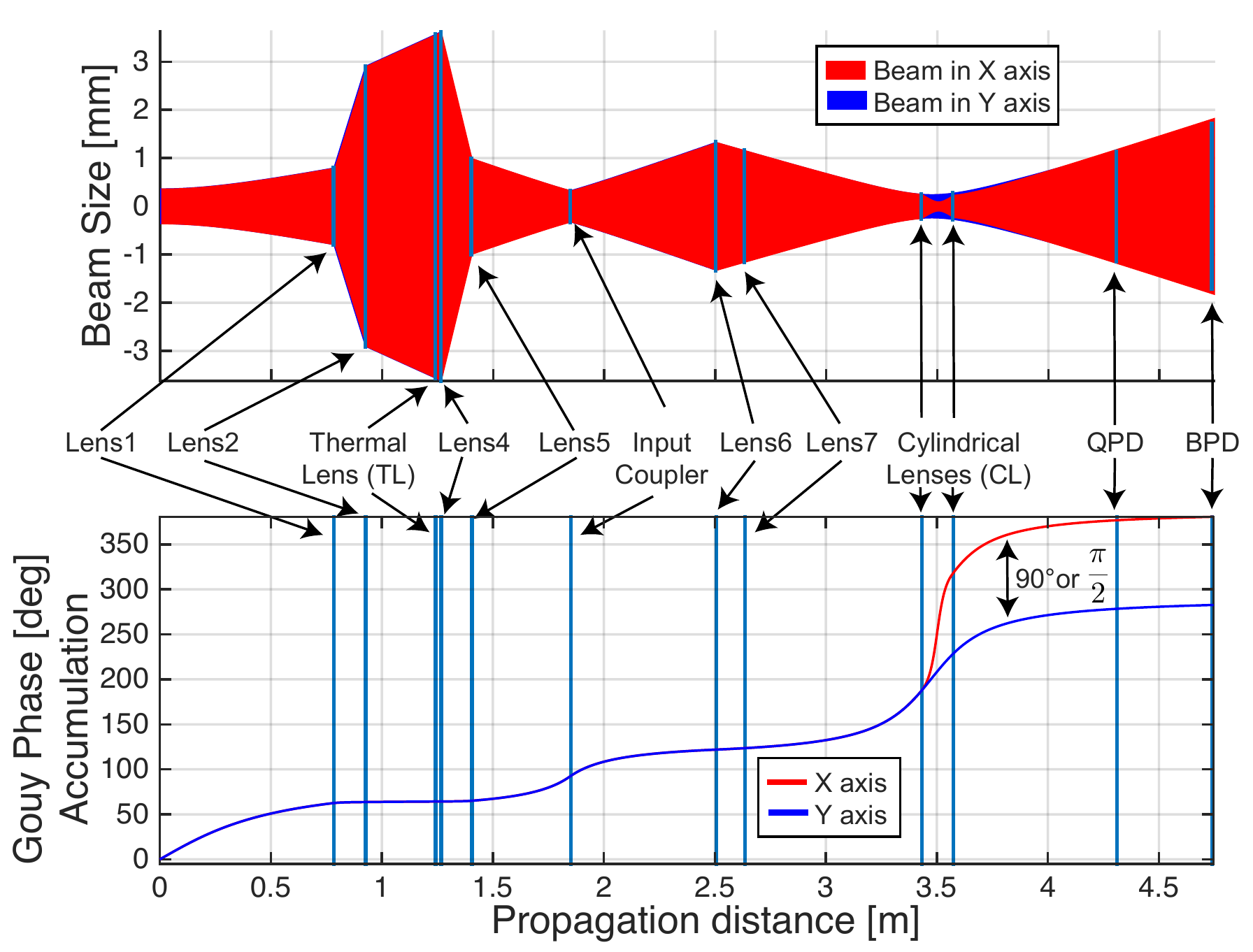}
\caption{\label{fig:TelescopeToWFS} Similar to FIG. \ref{fig:TelescopeToCav} the as built telescope layout is shown. Here the beam propagates from the mode cleaner to the two-mirror cavity input coupler and reflects instead of transmits. The reflected beam then propagates to beam focusing lenses and a cylindrical lens mode converter. The reflected beam terminates at either a QPD or BPD wavefront sensor at similar Gouy phases. The BPD is located on a separate beam with the same profile as the blue y-axis plot, see figure \ref{fig:ExperimentalSetup}. The vertical (Y) and horizontal (X) beam axis are shown. The only difference between the beams is seen at the $\tfrac{\pi}{2}$ mode converter. We show that one axis is focused while the other remains unchanged which adds a 90$^{\circ}$ Gouy phase difference between the axes. }
\end{figure}

\begin{table}
\centering
\begin{tabular}{|r|c|r|r|r|}
\hline
      & $f$ [m] & $d$ [m] & Gouy phase & Beam Size [$\mu$ m] \\
\hline
PMC   & N/A     & 0.0000 &   0$^{\circ}$ &  371 \\
Lens1 & -0.574  & 0.7805 &  62$^{\circ}$ &  803 \\
Lens2 & +2.291  & 0.9230 &  63$^{\circ}$ & 2914 \\
TL    & $-\inf$ to -10 & 1.2375 & 64$^{\circ}$ & 3572 \\
Lens4 & +1.719  & 1.2630 &  64$^{\circ}$ & 3625 \\
Lens5 & -0.574  & 1.4012 &  65$^{\circ}$ & 1000 \\
IC    & 0.33 m RoC  & 1.8462 &  92$^{\circ}$ &  333 \\
FP    & N/A     & 2.0102 & 137$^{\circ}$ &  236 \\
OC    & 0.33 m RoC  & 2.1742 & 182$^{\circ}$ &  333 \\
Lens6 & 0.45767 & 2.5022 & 122$^{\circ}$ & 1335 \\
Lens7 & $\inf$  & 2.6302 & 123$^{\circ}$ & 1166 \\
CL1   & +0.100  & 3.4274 & 186$^{\circ}$ &  260 \\
CL2   & +0.100  & 3.5688 & 227$^{\circ}$, 317$^{\circ}$ &  260 \\
QPD   & N/A     & 4.3058 & 278$^{\circ}$, 368$^{\circ}$ & 1145 \\
BPD   & N/A     & 4.7358 & 283$^{\circ}$, 373$^{\circ}$ & 1714 \\
\hline
\end{tabular}
\caption[Telescopic Design]{\label{fig:Table}
The as built parameters of the experimental setup seen in FIG. \ref{fig:ExperimentalSetup}, \ref{fig:TelescopeToCav}, \ref{fig:TelescopeToWFS} are listed. 
There are three paths to note: PMC through OC for the cavity; PMC to OC then reflected to Lens6, Lens7 and finaly the BPD; and PMC to OC then reflected to Lens6, Lens7, CL1, CL2, and finally to the QPD.
This table is used to obtain the Gouy phase difference between: the actuator and the cavity; the actuator and the QPD/BPD; and the cavity and the QPD/BPD.}
\end{table}


The experimental setup is very similar to our computer simulation described in appendix \ref{computersimulation} and in FIG. \ref{fig:FINESSELAYOUT}. 
Though both model and experiment conclude that a mode converter, paired with QPDs, is equivalent to the use of BPDs there are a few subtle differences.
The model uses four wavefront sensors. BPD2 and QPD2 are placed at an effective $0^{\circ}$ Gouy phase from the actuator. The second set, BPD1 and QPD1, are placed at an effective $45^{\circ}$ Gouy phase from the actuator.
In our experimental demonstration we place one BPD at \textcolor{black}{$283^{\circ}$} Gouy phase, which is an effective $39^{\circ}$ from our actuator after phase wrapping.
Additionally, we place one QPD at \textcolor{black}{$278^{\circ}$} Gouy phase, which after phase wrapping is at an effective \textcolor{black}{$34^{\circ}$} from our actuator.
Note that the Gouy phases for the QPDs are reported with respect to the non-focusing axis of the cylindrical lenses.
The Gouy phase along the cylindrical lens focusing axis is an additional $90^{\circ}$.
Also, in the model we changed the input beam complex beam parameter to simulate either waist size or waist location only.
In practice, our lens actuator caused a change in both waist size and waist location at the same time.

\subsection{Thermal Lens Actuator Telescope}

The thermal lens actuating telescope is composed of the first five lenses noted in FIG.\ref{fig:TelescopeToCav}, FIG.\ref{fig:TelescopeToWFS}, and Table 1. 
The first two lenses expand and collimate the beam into the thermal lens actuator while the last two mode-match into the optical cavity. 
A larger beam on the thermal lens will provide better actuation range.
The power overlap of the Gaussian beam before and after a thermal lens with focal length $f$ is given by 
\begin{equation}\label{LensoverlapFocal}
|I|^2 = 1 - (\tfrac{\pi w^2(z)}{2f\lambda})^2+O\left( \tfrac{w^8}{f^4\lambda^4} \right).
\end{equation}
Thus a large beam spot size is needed for effective actuation. Furthermore, an annually heated thermal lens with power $P_h$ produces a power overlap of
\begin{equation}\label{LensoverlapPower}
|I|^2 = 1 - \left( \tfrac{w(z)}{R_{optic}} \right)^4 \cdot \left( \tfrac{FOM\cdot P_h}{4\lambda} \right)^2
\end{equation}
where $FOM$ is obtained from \cite{arain2010adaptive}. This means that the two competing terms are the beam size and optic radius. 

\begin{figure}[h]
\centering
\includegraphics[width=.47\textwidth]{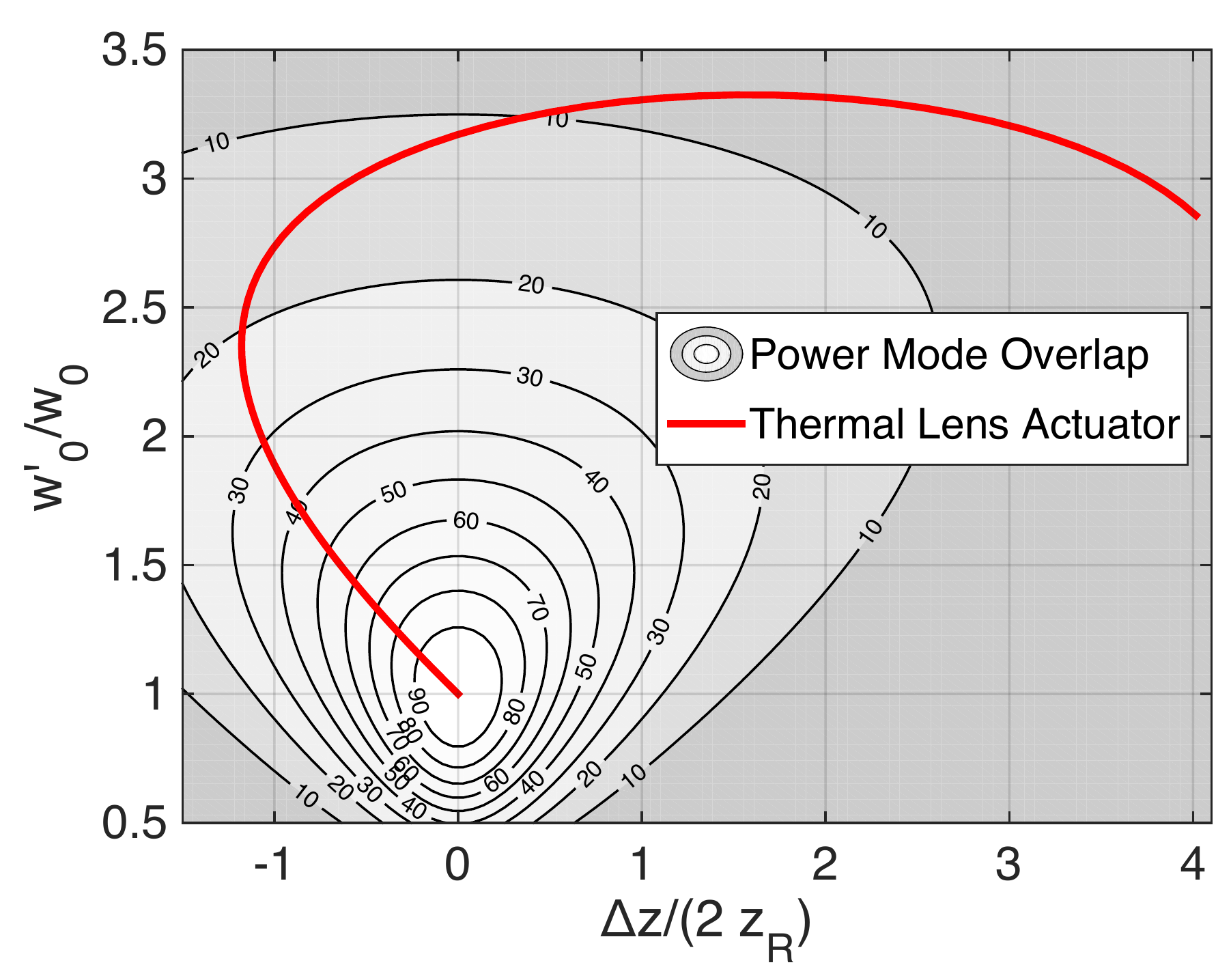}
\caption{\label{fig:ModeOverlapTLActuation} Power mode overlap $P = |\tfrac{2i\sqrt{z_{R1}z_{R2}}}{q_1-q_2^*}|^2$ between the optical cavity and the input beam is shown as the contour lines in percentage.
The thermal lens actuation path is seen in red.
As the thermal lens actuator changes the input beam into the cavity, the power mode overlap decreases.
Mode mismatching can reach well below 10\%.}
\end{figure}

Incorporating these principles into a design yielded a thermal lens actuating telescope that produced mode-matching between 100\% and slightly below 10\%. 
Though significant mode mismatch can be generated, wavefront sensors are best suited for measuring small amounts of misalignment or mode mismatch. 
This means that for relatively low input heating power, less than 5 watts, our thermal lens actuator telescope could measurably mismatch the beam into the optical cavity. 
The thermal lens actuation is further explained with FIG. \ref{fig:ModeOverlapTLActuation}.

In addition to mode mismatching, this thermal lens actuator also had the capability to create pitch and yaw misalignment. This was used to verify the preservation of alignment wavefront sensing.

%
%
%

\subsection{Wavefront Sensor Calibration}
\label{calibrationsec}
In this subsection we discuss how we calibrated the wavefront sensors and thermal lens actuator.
The field mode mismatch $\epsilon=(q'-q)/(q-q^*)$, generated by our actuator, is is ultimately converted to digital counts (cts) in the data acquisition system as follows.

The power mode mismatch $|\epsilon|^2$ was monitored via a DC photodiode in the transmission of our optical cavity.
The power drop percentage is proportional to the power mode mismatch $|\epsilon|^2$ as described by Anderson \cite{anderson1984alignment}.
Our thermal lens actuator was set up so that we could degrade the optical cavity mode matching from 100\% to just below 10\% as seen in FIG \ref{fig:ModeOverlapTLActuation}.
Though we had a wide range for mode matching, we chose to induce between 100\% and 91\% mode matching or 9\% mode mismatch. 

We next calculated from first principles the expected reflected RF power due to mode mismatch.
As stated in equation \ref{eq:I} from the appendix a certain amount of mode mismatch $\epsilon$ will induce the following reflected optical power in watts peak for the quadrant photodiode
\begin{equation}\label{powerwattspeak}
P_{\mbox{watts peak QPD}} = 4\Psi_{SSB}\Psi_{C}\Im{(\epsilon e^{i\phi_{GP}})}2\pi^{-1}
\end{equation}
and the following reflected optical power for the bullseye photodiode
\begin{equation}
P_{\mbox{watts peak BPD}} = 4\Psi_{SSB}\Psi_{C}\Im{(\epsilon e^{i\phi_{GP}})}2e^{-1}.
\end{equation}
Note that $\Psi_{C}$ is the carrier field extracted from directly measuring the optical cavity transmitted power $\Psi_{C}=\sqrt{P_{cavtrans}}$.
It should also be noted that $\Psi_{SSB}$ is the single-sideband field back-calculated from measured cavity transmitted power, cavity input power, cavity mirror measured transmissivity, and also includes a 0.95\% intra-cavity loss term.
The Gouy phase between the actuator and sensors $\Delta\phi_{G}$ can be read from the telescope table \ref{fig:Table} above for both the BPD and QPD.
The Gouy phase separation between the BPD sensor and the thermal lens actuator is $39^{\circ}$ while the Gouy phase seperation between the QPD sensor and the thermal lens actuator is $34^{\circ}$.

We can now compare the RF power in watts peak calculated from first principles to the RF power measured from calibrated electronics. 
The reflected beam first travels through several beam splitters which attenuate the beam by a factor of $A_{BPD}=.300$ for the bullseye and $A_{QPD}=.119$ for the quadrant.
The optical power is then converted to current at the photodiode.
All the electronics were calibrated by injecting voltage signals and measuring the output.
The response of the quadrant photodiode is 0.03 Amps/Watt at 1064 nm wavelength and has a transimpedance 10,000 Volts/Amp.
The response of the bullseye photodiode is 0.20 Amps/Watt at 1064 nm wavelength and has a transimpedance of 7,100 Volts/Amp.
These RF voltages are then demodulated with our LIGO-built wavefront sensing electronic crate.
The wavefront sensing crate demodulates the RF signal and contributes a factor of 6.7 gain. 
This gain was measured by injecting a 25 MHz sine wave at $12.7 \mbox{mV}$ peak-to-peak.
The demodulated signal was not constantly in phase so a 200 mHz wave at $190 \mbox{mV}$ peak-to-peak was observed.
If the injection was perfectly in phase we would see a DC voltage of $190 \mbox{mV}_{pp}$/2=85 mV.
From this we calculate the factor of 6.7 gain by $6.7=85\mbox{mV}/12.7\mbox{mV}_{pp}$.
Now the demodulated signals are relatively low frequency and are sent to the digital system.
The digital system has low pass filters, but do not alter the demodulated signals.
We injected a known voltage into the digital data acquisition system and obtained a conversion of $\frac{1 Volt}{1326 cts}$.
Combining the beam splitter attenuation and all electronic gains leads to a direct conversion from cts to radio frequency optical watts peak at 25 MHz.

For the quandrant photodiode we have
\begin{center}
$P_{\mbox{watts peak QPD}}\cdot A_{QPD}  = $
$\mbox{cts}\cdot\frac{1 V}{1326C}\cdot6.7\cdot\frac{1 A}{10,000 V}\cdot\frac{1W}{0.03A}$
\end{center}
and for the bullseye photodiode we have
\begin{center}
$P_{\mbox{watts peak BPD}}\cdot A_{QPD}  = $
$\mbox{cts}\cdot\frac{1 V}{1326C}\cdot6.7\cdot\frac{1 A}{7,100 V}\cdot\frac{1W}{0.2A}.$
\end{center}
We compress this whole calibration into a term $C_{Q}=\frac{1 V}{1326C}\cdot6.7\cdot\frac{1 A}{10,000 V}\cdot\frac{1W}{0.03A}/A_{QPD}$ for the quadrant photodiode and similarly for the bullseye photodiode $C_{B}=\frac{1 V}{1326C}\cdot6.7\cdot\frac{1 A}{7,100 V}\cdot\frac{1W}{0.2A}/A_{BPD}$.




We then solve for mode mismatch $\epsilon$ and have a fully calibrated expression in terms of counts (cts).
\begin{equation}
\epsilon_Q=\frac{cts_Q\cdot C_Q}{4 A_Q \Psi_S \Psi_C \frac{2}{\pi} (-\cos{(2\pi \Delta \phi_G}))} 
\end{equation}

\begin{equation}
\epsilon_B=\frac{cts_B\cdot C_B}{4 A_B \Psi_S \Psi_C 2e^{-1} (-\cos{(2\pi \Delta \phi_G} ))} 
\end{equation}

\subsection{Experimental Results}
\begin{figure}[h]
\centering
\includegraphics[width=.5\textwidth]{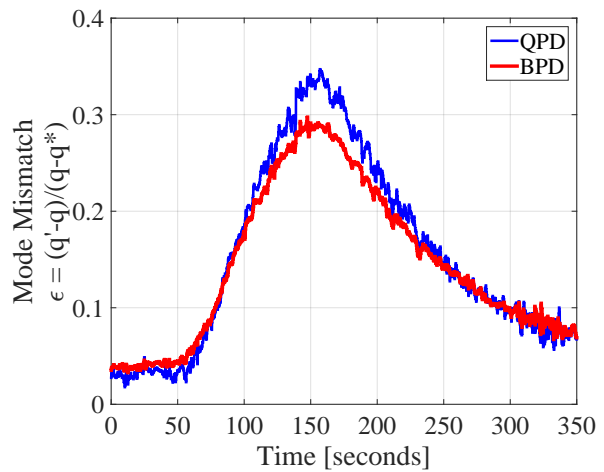}
\caption{\label{fig:ExperimentResultsToModeMatching} Measured mode-mismatch response is shown as 9\% mode mismatch is manually induced using the calibrated thermal lens actuator. The thermal lens actuator is equally heated radially thus ensuring only mode mismatch was induced. Counts from our digital data acquisition system are converted into mode mismatch $\epsilon$ as stated in \ref{calibrationsec}. From first principles our calibration also yields a $|\epsilon|^2=9\%$ mode-mismatch at the maximum value on this plot. Finally we also observed a 9\% drop in cavity transmitted power.}
\end{figure}

The results show good agreement between the bullseye photodiode (BPD) and the mode-converted quadrant photodiode (QPD) as seen in FIG. \ref{fig:ExperimentResultsToModeMatching}.
Additionally, both QPD and BPD measured 9\% mode mismatch which is consistent with the 9\% mode mismatch induced by the thermal lens actuator.
Note that the photodiode placement was chosen to reduce the number of lenses needed and to be relatively far away from a beam focal point, such that the beam size could easily match the photodiode size. This however resulted in a sub-optimal readout Gouy phase choice (QPD: $34^{\circ} + n \cdot 90^{\circ}$, BPD: $39^{\circ} + n \cdot 90^{\circ}$, where $45^{\circ} + n \cdot 90^{\circ}$ would be orthogonal.) 
Though this was a sub-optimal design choice, our results still clearly demonstrate the robustness of the heterodyne detection scheme.  
An ideal effective Gouy phase accumulation between an actuator and sensor should be a multiple of $90^{\circ}$.


The small discrepancy between the amplitude of the BPD and QPD error signals in FIG. \ref{fig:ExperimentResultsToModeMatching} may be due to the in phase (I) and quadrature phase (Q) manual tuning.
In the tuning we manually adjust the gain until the quadrature signal is extinguished.
However, the quadrature signal does not always go exactly to zero.  
The computer simulation in the appendix is better suited for comparing ideal BPD and ideal mode converted QPD error signals. 
It should be noted that even the idealized simulation contains some gain discrepancy which is due to the geometry of the photodiodes.


\section{Conclusion}

We theoretically derived the mode-matching error signal for bullseye photodiode (BPD) and mode-converted quadrant photodiode (QPD) wave front sensing. We showed that a mode-converted quadrant photodiode preserves the ability to measure alignment whilst enabling the ability to measure mode-match. We proposed a sensing scheme usable by any heterodyne optical setup directed towards Advanced LIGO, and experimentally demonstrated a side-by-side comparison of bullseye photodiode and mode-converted quadrant photodiode sensing.
We should also point out that using a mode-converted quadrant photodiode shifts the difficulties in setting up a bullseye photodiode Gouy phase telescope with a specific beam size to the placement of the mode converter lenses, which is much easier to fine adjust. 

We conclude that this mode-converter-based sensing scheme could yield a non-invasive, inexpensive mode-matching upgrade to terrestrial gravitational-wave detectors such as Advanced LIGO, Advanced Virgo and KAGRA. All RF quadrant photodiodes used for interferometer alignment in those detectors could be upgraded by redesigning their respective Gouy phase telescopes to include cylindrical lenses.


\section{Appendix}

\subsection{Hermite-Gaussian modes with two complex q parameters}
\label{sec:HG2p}
\subsubsection{Complex Beam parameters}
The complex beam parameter of a Gaussian beam with Rayleigh range $z_R$, at a distance $z$ from its waist, is defined as
\begin{equation}
q= z+ i z_R \,\,\, .
\end{equation}
Beam size $w$ and phase front radius of curvature $R$ are then given by
\begin{equation}
\frac{1}{q}= \frac{1}{R} -  i \frac{\lambda}{\pi w^2}\,\,\, ,
\end{equation}
where $\lambda=2 \pi / k$ is the wave length of the light.
It allows expressing the Gaussian beam in a simple form:
\begin{equation}
\Psi(x,y,q)=A(x,y,q) e^{-i k z} 
\end{equation}
\begin{equation}
\label{myequation4}
A(x,y,q) = \frac{A}{q} e^{-i k \frac{x^2+y^2}{2 q}} 
\end{equation}
where A is a complex constant (amplitude). It can be helpful to introduce the field amplitude on the optical axis, $\psi=A/q$, which now evolves along the z-axis due to the Gouy phase evolution, but is unaffected when passing through a thin lens. Thus, for any given location on the optical axis z, the Gaussian beam is completely described by the two complex parameters $\psi$ and $q$.
The main advantage of this formalism becomes apparent when using ray-transfer matrices $M$ defined in geometric optics (e.g. Saleh, Teich) to represent the action of a full optical system. The two complex parameters ($q_f$,$\psi_f$) after the system are given in terms of the initial parameters ($q_i$,$\psi_i$) by
\begin{equation}
\label{myequation5}
M \begin{pmatrix} \frac{1}{\psi_i} \\ \\ \frac{1}{\psi_i q_i} \end{pmatrix} = \begin{pmatrix} \frac{1}{\psi_f} \\ \\ \frac{1}{\psi_f q_f}\end{pmatrix} \,\,\,,
\end{equation}
and the change of the Gouy phase through the system, $\Delta \phi$, is given by
\begin{equation}
\label{myequation6}
e^{i \Delta \phi} = \sqrt{\frac{\psi_f}{\psi_f^*} \frac{\psi_i^*}{\psi_i}  } \,\,\, .
\end{equation}
This expression is consistent with the usual definition of local Gouy phase for a Gaussian beam as $\phi = \arctan z/z_R$,
but preserves the Gouy phase when propagating through a lens. To prove expressions \ref{myequation5} and \ref{myequation6} it is sufficient to verify them for a pure free-space propagation and a pure lens.

If we now introduce astigmatism, either intensionally with cylindrical lenses or accidentally through imperfections, cylindrical symmetry around the beam axis will be lost. As long as we introduce this astigmatism along a pre-determined axis (say the x-axis), we can simply proceed by introducing separate q-parameters for the x- and y-axis, $q_x$ and $q_y$. Since ray-transfer matrices are introduced with only 1 transverse axis, the propagation of $q_x$ and $q_y$ is done with ray-transfer matrices defined for the corresponding transverse axis. Thus we now have a separately-defined beam size $w_x$, $w_y$, phase front radius of curvature $R_x$, $R_y$, Rayleigh range $z_{Rx}$, ${z_Ry}$ and Gouy phase $\phi_x$ and $\phi_y$ for each of the two transverse directions. The corresponding fundamental Gaussian beam is given by
\begin{equation}
\label{eq:fundamentalXY0}
\Psi(x,y,q_x,q_y)=A(x,y,q_x,q_y) e^{-i k z} 
\end{equation}
\begin{equation}
\label{eq:fundamentalXY}
A(x,y,q_x,q_y) = \frac{A}{\sqrt{q_x q_y}} e^{-i k \frac{x^2}{2 q_x}}  e^{-i k \frac{y^2}{2 q_y}} 
\end{equation}
where A is again a complex amplitude.
Next we introduce the Hermite-Gaussian basis set corresponding to the fundamental Gaussian beam. In the literature this is typically done only relative to a single q-parameter, but it directly generalizes to the case with separate $q_x$ and $q_y$ parameters:
\begin{equation}
\label{eqnHGdef1}
\Psi_{nm}(x,y,q_x,q_y)=A_{nm}(x,y,q_x,q_y) e^{-i k z} 
\end{equation}
\begin{equation}
A_{nm}(x,y,q_x,q_y)=N A_{n}(x,q_x) A_{m}(y,q_y)
\end{equation}
\begin{equation}
\label{eq:HGXY}
A_p(\xi,q_\xi) =e^{i p \phi_{\xi}}
\sqrt{\frac{1}{2^p p!} \psi_\xi}
\,\,\, H_p(\sqrt{2} \frac{\xi}{w_\xi} )
\,\,\, e^{-i k \frac{\xi^2}{2 q_\xi}} 
\end{equation}
\begin{equation}
\psi_\xi = {\sqrt{\frac{2}{\pi}} } \,\,\, \frac{e^{i\phi_{\xi}}}{{w_{\xi}}}
= \sqrt{\frac{2 z_R}{\lambda}} \frac{i}{q_\xi}
\end{equation}
\begin{equation}
\label{eqnHdef}
H_0(\eta) = 1 \,\,\,,\,\,\, H_{p+1}(\eta) = 2 \eta H_p(\eta) - \frac{d}{d \eta}H_p(\eta)
\end{equation}
Here, we redefined the overall amplitude $N$ such that the total power P in a mode is simply given by $P = \int |\Psi_{nm}|^2 dx dy =|N|^2$. That equations \ref{eq:fundamentalXY0} and \ref{eq:fundamentalXY} are of the same form as equations \ref{eqnHGdef1} to \ref{eqnHdef} can be seen by using the identity $i z_R / q = e^{i \phi} w_0 / w$. 
Furthermore we defined $\psi_\xi$ in analog to the field amplitude $\psi$ introduced after equation \ref{myequation4}, that is the field amplitude on the optical axis of the fundamental mode. It thus evolves, together with $q_\xi$,  according to equations \ref{myequation5} and \ref{myequation6}. Note though that there is an extra Gouy phase term for the higher order modes that is explicitly excluded from the definition of $\psi_\xi$. As a result, the overall Gouy phase evolution of $\Psi_{nm}(x,y,q_x,q_y)$ is proportional to $e^{i (n + 1/2) \phi_{x} + i (m + 1/2) \phi_{y}}$.

As expected, these modes still solve the paraxial Helmholtz equation
\begin{equation}
(\bigtriangleup_T -2 i k \frac{\partial}{\partial z}) A_{nm}(x,y,q_x,q_y) = 0
\end{equation}
exactly.
Finally, in the main text we use the simplified bra-ket notation for readability:
\begin{equation}
|HG_{nm}\rangle =|\Psi_{nm}(x,y,q_x,q_y)\rangle.
\end{equation}
Specializing to the non-astigmatic $q_x=q_y$ we also use the two identities 
\begin{equation}
\label{eqLGdef}
|LG_{01}\rangle =\frac{1}{\sqrt{2}}|HG_{20} \rangle + \frac{1}{\sqrt{2}}|HG_{02} \rangle,
\end{equation}
\begin{equation}
\label{eqHGrotdef}
|HG_{11}^{45^{\circ}{\rm rot}}\rangle =\frac{1}{\sqrt{2}}|HG_{20} \rangle -\frac{1}{\sqrt{2}}|HG_{02} \rangle.
\end{equation}
Equation \ref{eqLGdef} relates the Hermite-Gaussian basis to the Laguerre-Gaussian basis (see e.g. \cite{o2000mode}) , while equation \ref{eqHGrotdef} directly follows from equations \ref{eqnHGdef1} to \ref{eqnHdef} under a $45^{\circ}$ rotation around the beam axis.

\subsubsection{Design of the $\frac{\pi}{2}$ mode-converter}
\label{DesignofMC}

Equations \ref{eqLGdef} and \ref{eqHGrotdef} highlight that the key requirements for a mode-converter capable of converting a $|LG_{01}\rangle$ into a $|HG_{11}^{45^{\circ}{\rm rot}}\rangle$ mode: We need a difference of $\pi$ in phase evolution between the two 2nd order modes $|HG_{20}\rangle$ and $|HG_{02}\rangle$, leading to a relative sign flip. We thus require a telescope consisting of at least two cylindrical lenses that
\begin{enumerate}
\item has a x-Gouy phase $\Delta \phi_x$ and y-Gouy phase $\Delta \phi_y$ evolution that differs by exactly $\frac{\pi}{2}$ between the first and last cylindrical lens ($\Delta \phi_x - \Delta \phi_y = \frac{\pi}{2}$), and
\item again matches the x- and y- Gaussian parameters $q_x$ and $q_y$ after the last cylindrical lens.
Note that technically the quadrant photo detector (QPD) could be placed at the location of, and instead of the last cylindrical lens. But that would make any further downstream adjustment of the sensing Gouy phase of the QPD impossible.
\end{enumerate}

While there are an infinite number of solutions that fit conditions 1) and 2) above, there is only one symmetric solution with two cylindrical lenses with the same focal length $f$ and the waist exactly in the middle between the two lenses. For this symmetric case, condition 2) requires the x- and y- beam size to be identical at the lenses:
\begin{equation}
\Im \left( \frac{1}{q_{x}}  - \frac{1}{q_{y}} \right)= 
\Im \left( \frac{1}{\frac{d}{2}+i z_{Rx}}  - \frac{1}{\frac{d}{2}+i z_{Ry}} \right)
=0,
\end{equation}
where $d$ is the separation between the lenses, $z_{Rx}$, $z_{Ry}$ are the Rayleigh ranges for the x- and y- Gaussian beam profile, and $\Im$ denotes the imaginary part. Excluding the trivial solution $z_{Rx}=z_{Ry}$, this implies the condition
\begin{equation}
\frac{d}{2 z_{Rx}} \cdot \frac{d}{2 z_{Ry}} = \tan{\frac{\Delta \phi_x}{2}} \cdot \tan{\frac{\Delta \phi_y}{2}} =1.
\end{equation}
This is equivalent to 
\begin{equation}
\cos{\frac{\Delta \phi_x + \Delta \phi_y}{2}} =0. 
\end{equation}
Using $\Delta \phi_x - \Delta \phi_y = \frac{\pi}{2}$ from condition 1., we thus find
\begin{equation}
\Delta \phi_x = \frac{3 \pi}{4}\,\,\, , \,\,\, \Delta \phi_y = \frac{\pi}{4}.
\end{equation}
Finally, since $\tan{\frac{\pi}{8}} = \frac{1}{\sqrt{2} + 1}$ and $\tan{\frac{3 \pi}{8}} = \frac{1}{\sqrt{2} - 1}$, we get for the cylindrical focal length $f$ of both lenses and the lens separation $d$
\begin{equation}
\label{eq:MCFocalLength}
f=\frac{z_0}{1+\frac{1}{\sqrt{2}}} \,\,\, , \,\,\, d=\sqrt{2} f ,
\end{equation}
where $z_0=z_{Ry}=\frac{\pi w_0^2}{\lambda}$ is the Rayleigh range of the incoming beam (no lens in y-direction).

\subsection{Comparison to sensing with a bull's-eye detector}
\label{sec:comp}

We use the term bull's-eye photo-diode (BPD) for a photodiode with a center segment of radius $r$, and additional outer segments arranged in a ring around the central segment. Typically there are three outer segments to still get alignment information from the detector (see figure \ref{fig:Quadrant_Photodiode}, right side).

When sensing mode mismatch with a BPD, matching the center segment radius $r$ to the Gaussian Beam spot size $w$ via $w=\sqrt{2} r$ maximizes the mode-mismatch small signal sensing gain, because at that radius the $|LG_{01}\rangle$ mode has a node. 
However, for this choice we find that any residual length fringe deviation will couple directly into the mode-mismatch error signal because
\begin{equation}
\langle HG_{00}|BPD|HG_{00}\rangle = 1-2 \e^{-1} \approx 0.2642 \neq 0,
\end{equation}
where $BPD$ is equal to 1 on the central segment ($x^2+y^2<r$), and -1 on the outer segments ($x^2+y^2>r$).
This coupling can be reduced to zero by choosing $r'=w \sqrt{{0.5 \ln{2}}}$ as central segment radius, at the cost of some optical gain (see below). Either way though the BPD has to be matched in size to the Gaussian beam. This often makes adjusting the sensing Gouy phase of a BPD a bit awkward, since it is not possible to simply slide the detector across the optical axis. Furthermore, the amount of clipping on the bull's-eye photo-diode is set at the time of manufacturing by the size of the outer ring segments.

In contrast, a quadrant photo-diode (QPD) placed after a $\frac{\pi}{2}$ mode-converter has none of these beam size constraints. Instead, the reference beam size is set by the choice of the mode-converter through equation \ref{eq:MCFocalLength}, and can be changed by replacing the cylindrical lenses. The QPD can be moved freely to optimize the sensing Gouy phase and clipping, while any residual length fringe deviation does not couple to first order, since for a well-centered beam we find 
\begin{equation}
\langle HG_{00}|QPD|HG_{00}\rangle =  0.
\end{equation}
Here we chose $QPD={\rm sign}({x^2-y^2})$.

\subsection{Signal Gain for Sensing Mode-Mismatch}
\label{appSG}
Since we want to sense a mode-mismatched Gaussian beam $|HG_{00}^{q'}\rangle$ with beam parameter $q'$,
we can expand this beam in the unperturbed basis ($q$) as
\begin{equation}
\label{eq:expansion}
|HG_{00}^{q'}\rangle = \e^{-i \Im{\epsilon}}\sqrt{1-|\epsilon|^2} \, |HG_{00}^{q}\rangle  + \epsilon \, |LG_{01}^{q}\rangle + O(\epsilon^2),
\end{equation}
where $\Im$ denotes the imaginary part and $\epsilon$ encodes the waist size change $\Delta w_0$ and waist displacement $\Delta z$ of the Gaussian beam via 
\begin{equation}
\label{eq:epsilonDef}
\epsilon = \frac{q'-q}{q-q^*} = \frac{\Delta w_0}{w_0} - i*\frac{\Delta z}{2 z_R}
\end{equation}
Equation \ref{eq:expansion} includes enough $O(\epsilon^2)$ terms such that the power coupling is accurately given to 2nd order by
\begin{equation}
\label{eq:epsilonPower}
|\langle\!HG_{00}^{q} | HG_{00}^{q'}\rangle |^2= 1-|\epsilon|^2 + O(\epsilon^3).
\end{equation}

To calculate the small signal gain for a mode-sensing scheme we need the matrix element
\begin{equation}
\label{eq:gammaBPD}
\gamma_{B}=\langle HG_{00}|BPD|LG_{01}\rangle = - 2 \e^{-1}  \e^{2 i \phi} \approx -0.7358 \, \e^{2 i \phi} ,
\end{equation}
where $\phi$ is the Gouy phase at the BPD. The minus sign is an artifact of the definition of Laguerre-Gaussian modes \cite{o2000mode}. Here the central element radius of the BPD is $r=w/\sqrt{2} $. For a BPD with central segment radius $r'=w \sqrt{{0.5 \ln{2}}}$ the numerical pre-factor drops to $-ln(2) \approx -0.6931$. See section \ref{sec:comp} for a discussion.

The equivalent matrix element for a QPD, after converting the $|LG_{01}\rangle$ mode into a $|HG_{11}^{45^{\circ}{\rm rot}}\rangle$ mode, is 
\begin{equation}
\label{eq:gammaQPD}
\gamma_{Q}=\langle HG_{00}|QPD|HG_{11}^{45^{\circ}{\rm rot}}\rangle = \frac{2}{\pi} \, \e^{2 i \phi} \approx 0.6366 \, \e^{2 i \phi} .
\end{equation}

If we use this approach to sense the matching of a cavity (beam parameter $q'$) to its input beam using the Pound-Drever-Hall (PDH) approach, we will use an up-front RF phase modulation (modulation index $\Gamma$) with a sideband frequency that is not resonant in the cavity.
The Gaussian beam reflected from this cavity has the structure
\begin{equation}
\label{eq:PsiOut}
|\Psi_{in}\rangle = |HG_{00}^{q'}\rangle_C  + \frac{i \Gamma}{2}|HG_{00}^{q}\rangle_{+} + \frac{i \Gamma}{2}|HG_{00}^{q}\rangle_{-} + O(\Gamma^2),
\end{equation}
where the indices $C$, $+$ and $-$ indicate carrier, upper and lower sideband.
We can sense this beam with either a BPD or a QPD behind a mode-converter, and demodulate the signal's I quadrature. We find in first order of $\Gamma$ and $\epsilon$
\begin{equation}
\label{eq:I}
I=P \Gamma \Im{(\gamma \epsilon)},
\end{equation}
where $P$ is the effective power on the photo diode - that is ignoring any power that does not contribute the RF signal, $\Gamma$ is the modulation index, $\Im$ denotes the imaginary part, $\gamma$ is the matrix from equation \ref{eq:gammaBPD} or \ref{eq:gammaQPD}, and $\epsilon$ is defined through equations \ref{eq:expansion}, \ref{eq:epsilonDef}, \ref{eq:epsilonPower}.
 
\begin{figure}[h]
\centering
\includegraphics[width=.5\textwidth]{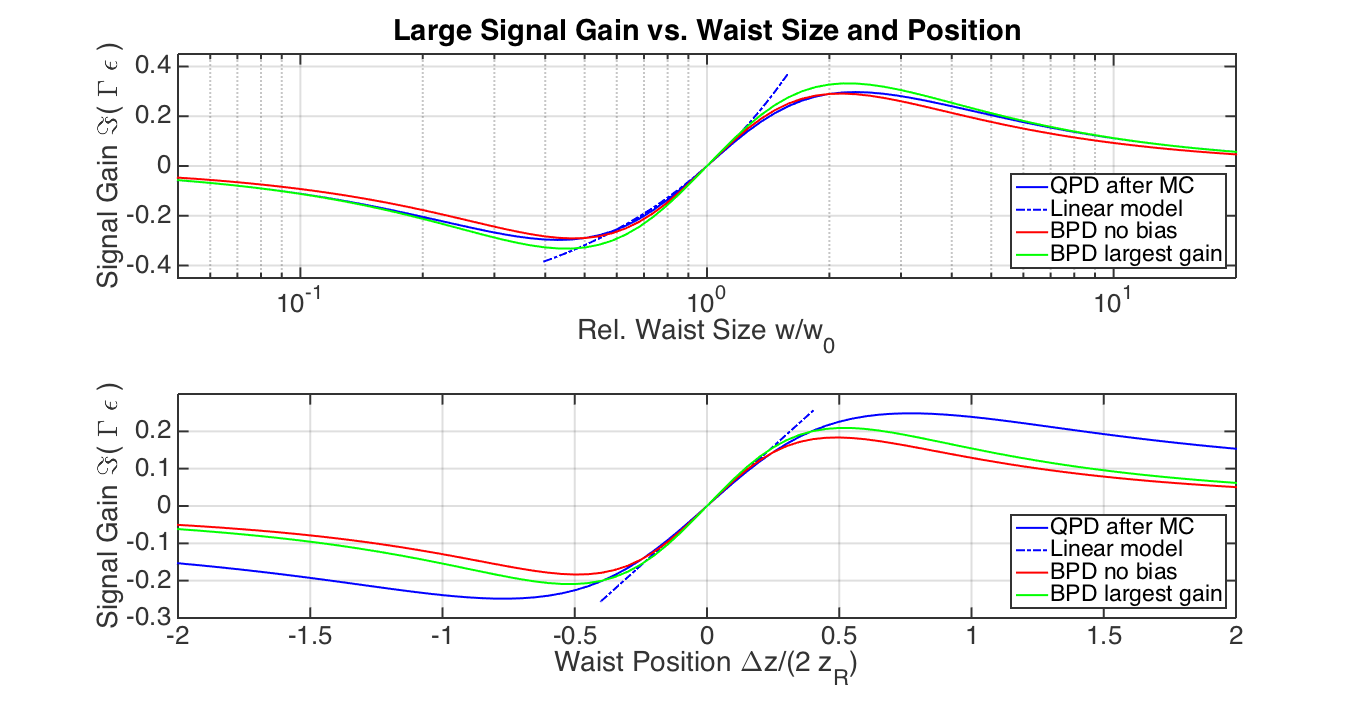}
\caption{\label{fig:LargeSignal} Large signal gain for mode sensing outside the linear regime. Plotted is the imaginary part of the matrix element in equation \ref{eq:I_lsig}, as a function of waist size (top) and waist location (bottom). For each plot the diode was placed in the optimal sensing Gouy phase. The solid traces blue, red and green are for a QPD placed after a mode-converter, a BPD with inner segment radius $r'=w \sqrt{{0.5 \ln{2}}}$ (no bias), and a BPD with inner segment radius $r=w \sqrt{0.5}$, in that order. All solid traces are calculated taking into account modes up or order $n,m=20$.The blue dash-dotted trace is the linear approximation from equation \ref{eq:gammaQPD} and \ref{eq:I}. Finally, The cavity is kept on resonance during the sweep - this affects the large signal behavior of all traces, as well as the small signal gain (slope) of the green trace in the lower plot (BPD largest gain). The small signal gains of the blue (QPD) and red (BPD no bias) are independent of any length offset.}
\end{figure}

For large mode deviations the power coupling from equation \ref{eq:epsilonPower} is given by the exact expression
\begin{equation}
\label{eq:epsilonPowerExact}
|\langle HG_{00}^{q} | HG_{00}^{q'}\rangle |^2= |\frac{2i \sqrt{(\Im{q'})(\Im{q})}}{q'-q^*} |^2,
\end{equation}
where $\Im$ denotes the imaginary part, and the sensing signal from equation \ref{eq:I} generalizes to 
\begin{equation}
\label{eq:I_lsig}
I=P \Gamma \Im{ (\langle HG_{00}^{q} | T^\dagger \,\, PD\,\, T |HG_{00}^{q'}\rangle)},
\end{equation}
where $PD$, is either the $BPD$ or the $QPD$. Here $T$ is the action of both mode-converter telescope (for the QPD) and Gouy phase telescope. Since we know the action of both telescopes on the two-parameter Hermite-Gaussian beams introduced in section \ref{sec:HG2p}, we can write the matrix element of equation \ref{eq:I_lsig} as
\begin{equation}
\label{eq:matrixElement_lsig_BPD}
\sum_{n,m}  \e^{i \phi (n+m)} \langle HG_{00}^{q} |  BPD\,  |HG_{nm}^{q}\rangle \langle HG_{nm}^{q} |HG_{00}^{q'}\rangle
\end{equation}
and
\begin{equation}
\label{eq:matrixElement_lsig_QPD}
\sum_{n,m}  i^{n} \e^{i \phi (n+m)} \langle HG_{00}^{q} |  QPD\,  |HG_{nm}^{q}\rangle \langle HG_{nm}^{q} |HG_{00}^{q'}\rangle .
\end{equation}
These expressions are plotted in figure \ref{fig:LargeSignal} with $\phi=0$ for waist location variations and $\phi=\pi/4$ for waist size variations, taking into account modes up to $n,m=20$. BPD and QPD have comparable, although not identical large signal gains.

\begin{figure}[h]
\centering
\includegraphics[width=.4\textwidth]{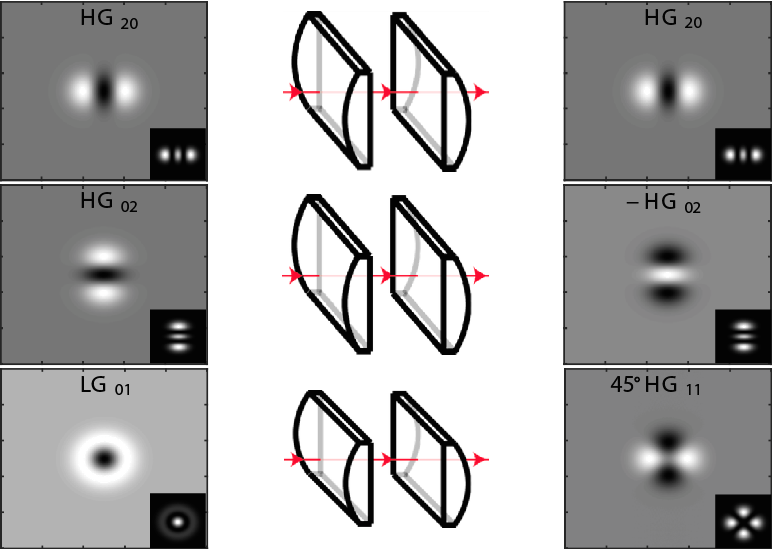}
\caption{\label{fig:ALLmodeconversions}MATLAB was used to model the $\frac{\pi}{2}$ mode converter using a Fourier optics representation of lenses. 
The cylindrical lenses both had a focal length of $f=0.1$m and were seperated by $f\sqrt{2}$.
The input beam waist was located half way between the cylindrical lenses and had a size of $ w_0 = (f\lambda(1+1/\sqrt{2})/\pi)^{1/2}$.
We propagate the $|HG_{02}\rangle$, $|HG_{20}\rangle$, and $|LG_{01}\rangle$ modes through the mode converting telescope. 
$HG$ modes oriented parallel or perpendicular to the lens focusing axis will experience no structural change in intensity profile (bottom right of each field image). $HG$ modes parallel to the lens focusing axis will get a sign flip in field. The $|LG_{01}\rangle$ mode converts into a $45^{\circ}$ rotated $|HG_{11}\rangle$ mode. We can also see that alignment $HG$ modes will be unaffected while mode mismatch $|LG_{01}\rangle$ modes will be perfectly converted into the $HG$ basis.}
\end{figure}

\subsection{Error Signal Model}
\label{computersimulation}
A computer simulation provided a convenient way for testing our prediction before performing the experiment. 
A combination of MATLAB and FINESSE \cite{FINESSEweb} was used to arrive at the mode mismatch error signal.
FINESSE uses ray transfer matrices while our MATLAB model uses the Fourier optic representation of lenses and beams.
FINESSE was previously used by Bond \cite{2016arXiv160601057B} to study optical cavity mode mismatch.  
That study served as a basis for comparison. 


\begin{figure}[h]
\centering
\includegraphics[width=.5\textwidth]{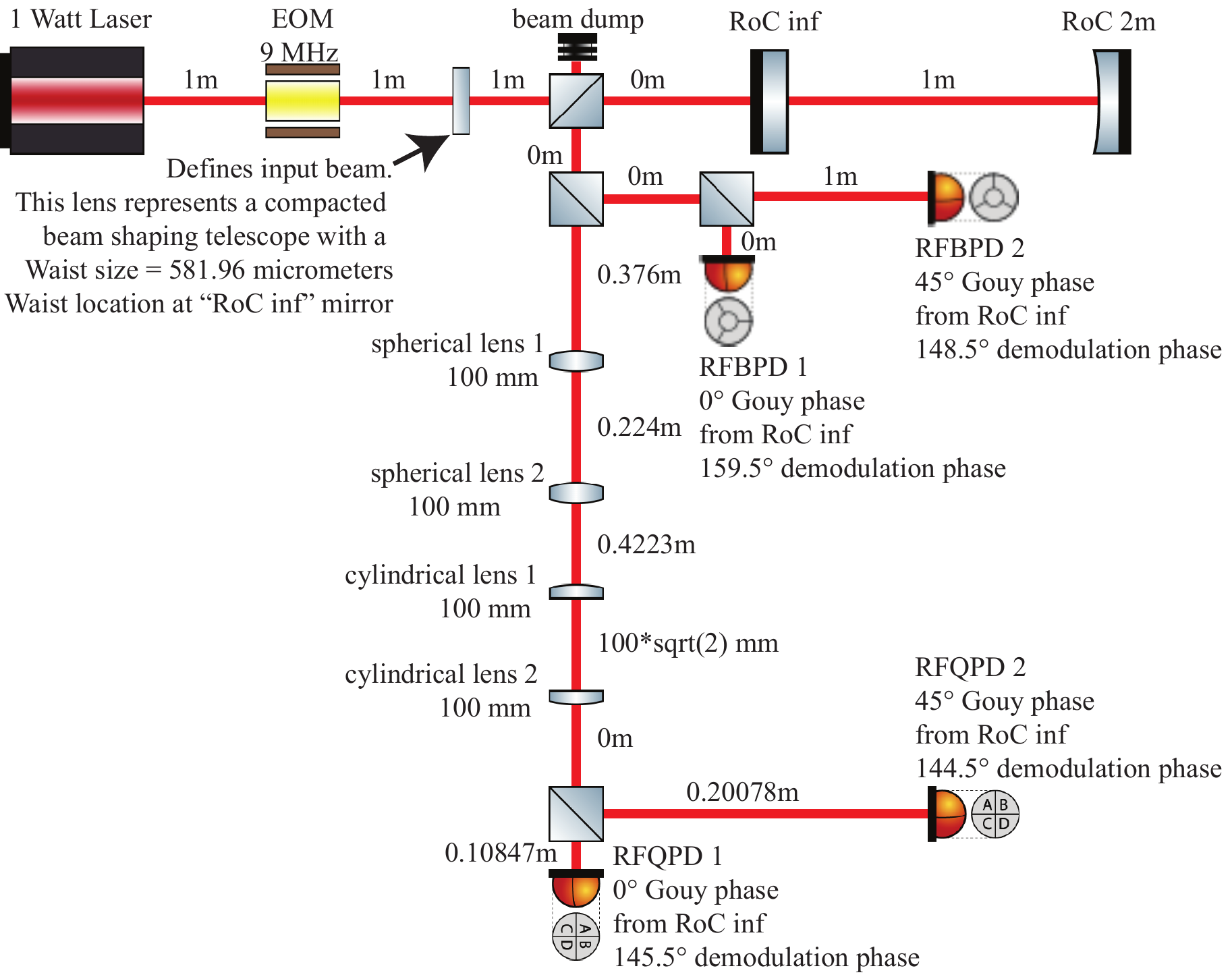}
\caption{\label{fig:FINESSELAYOUT} A 1 watt laser produces a beam at 1064 nanometer wave length. 
The beam passes through an Electro Optic Modulator (EOM) resonant at 9 MHz. 
The beam then passes through a beam splitter then into a hemispherical resonant optical cavity. 
The beam reflected from the cavity is then directed back to the beam splitter where now the reflected beam is directed to two paths. 
The first path contains two radio-frequency bullseye photodiodes (RFBPD) of varying radii. 
FINESSE automatically changes the bullseye photodetector size to match the beam incident on it. Secondly the beam passes through a beam shaping telescope then to a mode converter before finally arriving at two radio-frequency quadrant photodiodes (RFQPD). 
Each style of photodiode has one photodiode that measures the beam at $0^{\circ}$ Gouy phase and a second photodiode that measures the beam at $45^{\circ}$ Gouy phase. 
This Gouy phase separation is ideal for measuring both beam waist size and beam waist location.}
\end{figure}

The optical layout seen in FIG. \ref{fig:FINESSELAYOUT} was constructed to compare the error signals generated by bullseye photodiodes and quadrant photodiodes.
The input beam was varied in waist size and waist location.
This produced mode mismatching which was calculated in the reflected field.
Higher order modes beat against the fundamental sidebands yielding an error signal.
At this point, the field can be segmented and summed to reveal an error signal.
Measuring the reflected power at $0^{\circ}$ and $45^{\circ}$ Gouy phase will isolate both degrees of freedom. 

In the simulation, the bullseye photodiodes can measure mode mismatch at any Gouy phase since their sensing radius is automatically adjusted to fit the beam.
However, in practice the bullseye photodiodes are manufactured with one specific sensing radius so the incident beam needs to be shaped so that it not only fits, but also is at the correct Gouy phase. 

For quadrant photodiodes, the reflected field must first pass through beam shaping optics and then a cylindrical lens mode converter as seen in FIG. \ref{fig:FINESSELAYOUT}.
The field is then segmented into quadrants and the diagonals are summed and subtracted from the orthogonal diagonal. This can be better understood by seeing the error signal combination in FIG. \ref{fig:Quadrant_Photodiode}.

FIG. \ref{fig:ALLmodeconversions} shows the transverse electric field before and after it passes through a $\frac{\pi}{2}$ mode converter telescope.
The MATLAB model uses a heterodyne detection scheme to measure the beat between the fundamental sidebands and higher order mode mismatch modes
\cite{mueller2000determination}. The beam is phase-modulated at 25 MHz, and the photodiode output is demodulated with the same frequency. The cavity is kept locked on resonance.

\begin{figure}[h]
\centering
\includegraphics[width=.5\textwidth]{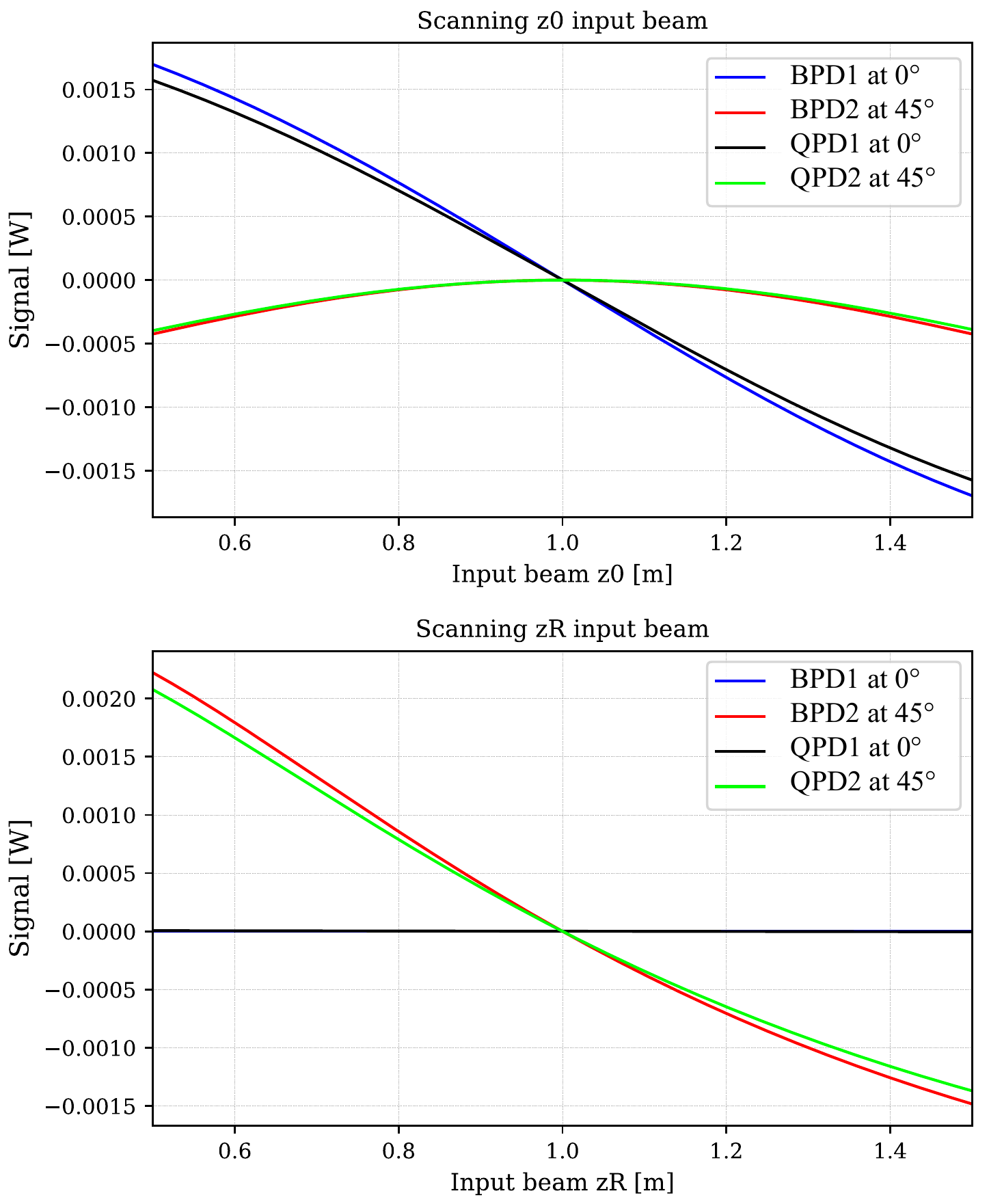}
\caption{\label{fig:FINESSEresults} Mode mismatch error signals generated by the FINESSE with MATLAB simulation. See appendix \ref{computersimulation} for more details.}
\end{figure}

The simulation results can be seen in FIG. \ref{fig:FINESSEresults} and show that we can generate a beam waist size and beam waist location error signal. 
We isolate beam waist size and beam waist position with both the bullseye and mode converted quadrant photodiode. 
The cavity input beam size is varied and results in the two error signal to the left. Notice that only the BPD and QPD placed at $45^{\circ}$ Gouy phase are sensitive to this kind of offset while the other two photodiodes see virtually no change. If instead we look at the second plot where the beam input beam waist position is shifted, we see that the opposite is true. Now the BPD and QPD placed at $0^{\circ}$ Gouy phase are sensitive to this kind of offset while the other photodiodes are not. This is the optimal placement for sensing mode mismatch. In practice we will want to also measure misalignment and thus we'll have to move the 2nd photodiode to somewhere between $45^{\circ}$ and $90^{\circ}$ Gouy phase, depending on the sensing noise requirements for alignment and mode-matching.
This simulation is a direct comparison between known methods of wave front sensing and our proposed scheme.

\section*{Acknowledgments}

We would like to thank the LIGO Scientific Collaboration for fruitful discussions held at the yearly LIGO VIRGO collaboration meetings. 
We would also like to acknowledge Antonio Perreca and Paul Fulda for providing simulation and hardware for the bullseye photodiodes. 
This work was supported by the National Science Foundation grant {PHY-1352511}. 
This document has been assigned the LIGO Laboratory document number {LIGO-P1900270}.
   
\bibliography{ModeConverter}
\bibliographystyle{unsrt}
\end{document}